# EUCARD MAGNET DEVELOPMENT

Gijs de Rijk, CERN, Geneva, Switzerland.

*Abstract*

The FP7-EuCARD work package 7 (WP7), "HFM: Superconducting High Field Magnets for higher luminosities and energies" is a collaboration between 12 European institutes and firms with the objective of developing high field magnet technology. WP7 foresees to construct a 13 T dipole with a 100 mm aperture, a $\Delta B = 6$ T high temperature superconductor (HTS) dipole insert, a superconducting HTS link and a superconducting helical undulator.

## EUCARD WP7 HIGH FIELD MAGNETS

The High Field Magnet work package is a collaboration between 10 institutes and 2 firms:
- CEA-Irfu Saclay, France (CEA)
- CERN, Genève, Switzerland (CERN)
- CNRS-Grenoble, France (CNRS)
- COLUMBUS, Genova, Italy (COLUMBUS)
- BHTS-Bruker, Hanau, Germany (BHTS)
- Karlsruhe Institute of Technology, Germany (KIT)
- INFN-LASA, Milano, Italy (INFN)
- Wroclaw Technical University, Poland (PWR)
- Southampton University, UK (SOTON)
- STFC-Daresbury, UK (STFC)
- Tampere Technical University, Finland (TUT)
- Université de Genève, Switzerland (UNIGE)

Besides a management task, the work package consists of 5 R&D tasks :
2. Support studies
3. High field model: 13 T, 100 mm bore ($Nb_3Sn$)
4. Very high field dipole insert (in HTS, up to $\Delta B = 6$ T)
5. High Tc superconducting link (powering links for the LHC)
6. Short period helical superconducting undulator (ILC $e^+$ source)

The duration is from April 2009 until April 2013. The total budget is 6.4 M€ from which 2.0 M€ is the EC contribution.

## HIGH FIELD MODEL

Several of the technologies used for $Nb_3Sn$ magnets (superconducting cable, insulation, coil design, support structures) were partly developed during the FP6-CARE-NED project. They are to be brought together and tested in a model dipole magnet. The aim of task 3 "High field model" is to design, build and test a 1.5 m long, 100 mm aperture dipole model with a design field of 13 T using $Nb_3Sn$ high current Rutherford cables.

The key component in a superconducting magnet is the conductor. In order to develop high field magnets it is essential to have a facility to tests the cables (not 'just' the strands) up to the maximum field and therefore this model will afterwards be used to upgrade the superconducting cable test facility FRESCA at CERN from 10 T to 13 T.

In Fig. 1 an overview is given of existing dipole magnets. In this figure, both magnets employed in accelerators and R&D models built to prospect high fields can be found. All the existing accelerators, which operate below 10 T, employ $\cos\Theta$ geometries with Nb-Ti conductors. Above 10 T both $\cos\Theta$ (D20 and MSUT) and block coil (HD1 and HD2) geometries were employed on models using $Nb_3Sn$ conductors. The proposed magnet (EuCARD-Fresca2) is at the top range of both field and aperture of all preceding projects. The design and construction of such a 13 T magnet with a 100 mm bore is thus an important challenge. To embark on such a project it is important to learn from existing HFM projects.

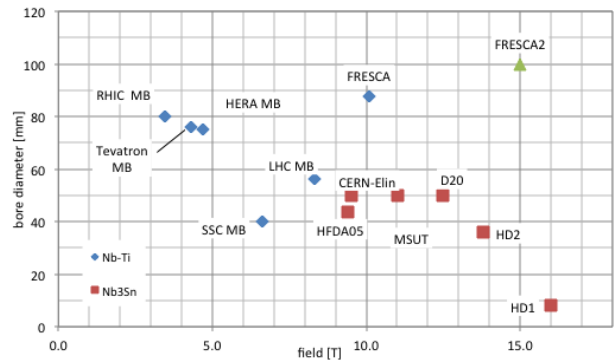

Figure 1: Field and bore diameter for a selection of superconducting magnets. For the magnets used in accelerators the fields are the real operational values. For model and prototype magnets these are the quench plateau values obtained during the tests.

During the first year of the project a study was made to compare potential coil geometries and a literature study was done on existing $Nb_3Sn$ magnets. In June 2010 the collaboration selected the block coil geometry for the EuCARD-Fresca2 magnet. This choice was backed by winding tests on the feasibility of the "flared-ends" which are needed for this type of coils (Fig. 2).

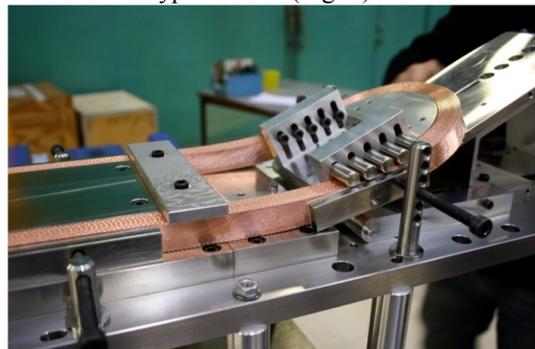

Figure 2: Winding tests at CEA for the block coil with flared ends

In Fig. 3 the choice and further development of the coil geometries in the design phase of this project can be seen.

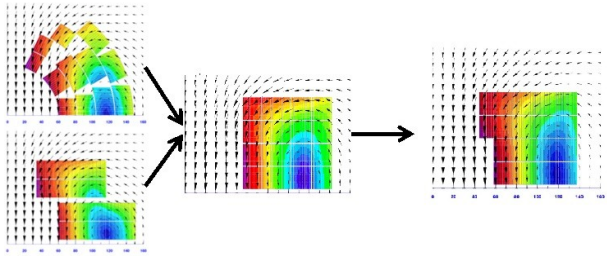

Figure 3: Development of the coil geometries during the design phase (one quarter shown).

For this magnet the conductor was selected taking into account previous developments in the CARE-NED program [1],[2] and by the LARP collaboration [3]. The cable has 40 strands of 1 mm diameter. Procurement of the strand has started and first prototype lengths have been delivered for tests.

The present layout consists of a coil with 2 double pancakes, the one close to the mid-plane has 36 + 36 turns and the outer one has 42 + 42 turns (see Fig 3 right hand picture). The picture of the magnet can be found in Fig. 4. The structure employs the shell-bladder and key system previously developed by LBNL [4]. At 13 T the magnet will operate at 82.5% of the load line at 4.2 K at a current of 10.6 kA given a degraded conductor performance of 1250 A/mm$^2$ at 15 T (this is 76.1% of the load line at 1.9 K). At this field the horizontal EM force is 16 MN/m and the stored energy is 3.6 MJ/m.

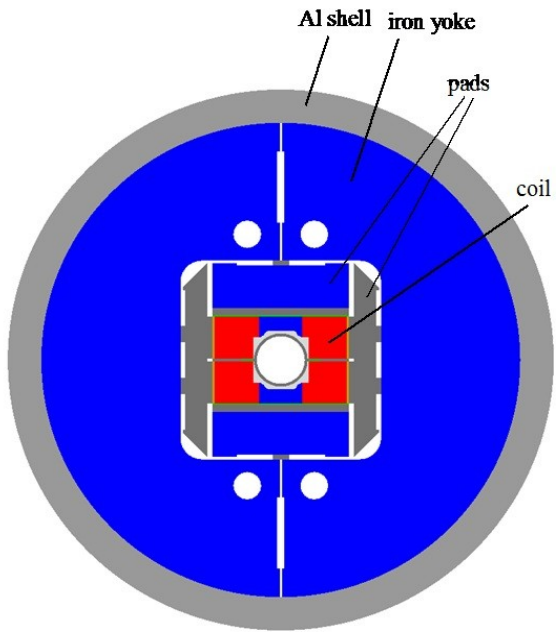

Figure 4: Schematic layout of the magnet

The main issues to be addressed for the 13 T dipole are:
- the conductor performance, quality and availability,
- the maximum field on coil,
- forces and stresses on the coil,
- the stored energy in the magnet,
- quench protection,
- the "makebility" of the coil and structure.

The structure applies nearly half of the pre-stress on the coil due to the differential shrinking between the shell and the yoke, the other half is applied at room temperature by inserting keys. The stress in the coil during the magnet lifecycle in one of the preliminary mechanical studies can be found in Fig. 5.

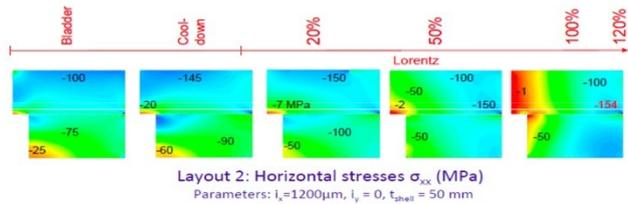

Figure 5: Coil stress during the magnet life-cycle

The flared ends of the coil imply that the cable is to be bend 'the hard way'; due to the natural elasticity of the cable the chosen bending radius of 700 mm is easy to execute. For comparison: in the HD2 magnet from LBNL the hard-way bending radius in the flared ends is 350 mm. In Fig. 6 a CAD picture of the coils in one pole can be found. In Fig. 7 a pre-design image of the ends of the magnet can be seen. Special attention will have to be paid to the design of the reaction tooling due to the combination of the thermal and reaction expansion of the conductor combined with the flared ends of the coils.

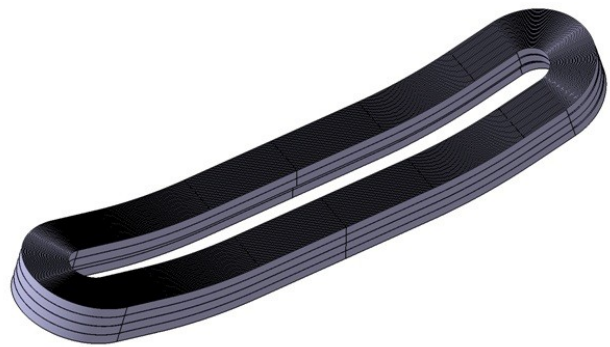

Figure 6: The coil of one pole, CAD image.

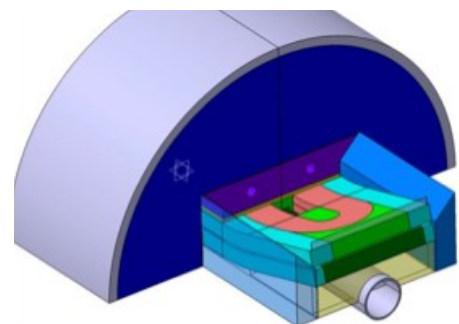

Figure 7: Pre-design image from the magnet ends.

The detailed design of the magnet was done in 2010 and a design review will be held on 20-21 January. The structure should be completed by the end of April 2011 and the mechanical behaviour in liquid nitrogen with dummy coils will be tested in the May-June 2011.

Component and tooling design will start end of 2010 and should be completed by April 2011. The critical path of the project is the conductor deliveries that are planned at regular intervals up to November 2011. The first double pancake coil with superconductor cable is planned to be ready by March 2012. Assembly of the magnet with a complete coils set will be started by end February 2013 followed by the magnet test in April 2013.

## VERY HIGH FIELD DIPOLE INSERT

Recent progress on High Temperature Superconductors like YBCO and BSCCO-2212 has shown good performance on the intrinsic current transport properties ($J_e > 400$ A/mm$^2$ at 4 K, B < 25 T). This should open the road to higher magnetic fields in the B = 20 T range interesting for HE-LHC. The aim this task is to design and fabricate an HTS very high field dipole insert (6-7 T), which can be installed inside the 13 T Nb$_3$Sn dipole of task 3 that will serve the role of the outer layer magnet. This is a very first attempt to approach 20 T in a dipole geometry. The development takes place in three steps. The first studies deal with the specification of several HTS conductors. This is to be completed by modelling work focused on stability and quench. The quench of HTS coils with very often occurring degradation is an identified issue. Due to the difficulty of making in one go a dipole insert coil of HTS conductor, several HTS solenoid insert coils will be made and tested in existing high field solenoid magnets. The experience that will be gained will be used to construct a dipole insert coil.

The main issues to be addressed for the dipole insert are:
- Jc of the HTS conductor: to reach 6 T we need an averaged Jc of ~ 300 A/mm$^2$;
- HTS coil fabrication;
- Electromagnetic forces in the range of 1000 t/m;
- Fixing into dipole;
- Coupling between dipole and insert, quenching either or both magnets.

The two candidate conductor types pose different strong and weak points:
1. BSCCO-2212 round strand:
- Good: cabling possibilities to reach high total currents.
- Poor: Critical heat treatment and weak mechanical performance.
2. YBCO tape:
- Good: Performance in Jc and stress (Fig. 8).
- Poor: cabling possibilities and difficult winding of coil ends.

Recent quench studies indicate that a quench of the 13 T magnet will quench the whole insert and thus a protection mechanism is inherently there for this case. Further quench studies are needed to cover all possible cases.

A first small solenoid made from YBCO has highlighted the issues to be solved: splice connections between the tapes need to be further developed and the fabrication process has to be optimized so as to avoid conductor degradation.

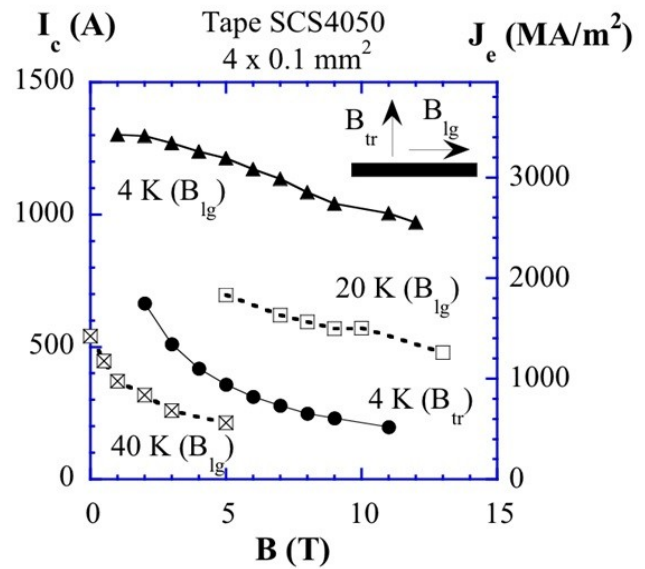

Figure 8: Measured critical current performance for a YBCO tape conductor sample tested in 2010

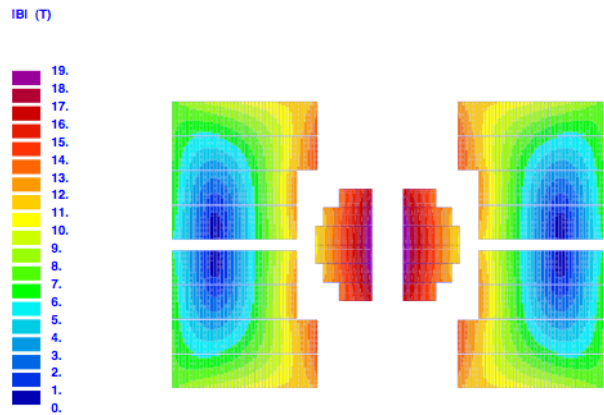

Figure 9: Field in the coils of the combined HTS insert and Nb$_3$Sn magnet

For the dipole insert a design was made using 12 mm wide YBCO coated conductor tape in a 'paired cable' geometry (see Fig. 9). The internal aperture is 20 mm in diameter. The current in a 12 mm tape is 610 A. The HTS dipole is located inside a 4 mm thick steel tube to contain the Lorentz forces (14 MN/m -16 MN/m) (in B=13 T from the outer dipole).

## HIGH TC SUPERCONDUCTING LINK

The interest of buses linking superconducting magnets made of HTS material was recognized already before the LHC startup. In one of the cleaning insertions this will be needed to replace a Nb-Ti superconducting link which will be at a thermal limit due radiation heating. Recently an additional problem has been identified with the radiation sensitivity of electronics, which renders the power convertors vulnerable. For running at high intensity and luminosity these problems are also felt in caverns close to the beam. To avoid the limitations imposed by these effects power convertors for the low beta insertion will have to be relocated in caverns far from the main tunnel or on the surface (see Fig. 10). For these type of solutions superconducting HTS links are needed to make efficient connections to the superconducting magnets using a minimum of space. The use of HTS enables operation at higher temperatures and offers a convenient gain in temperature margin during operation. In cases where space is limited and the radiation environment is harsh, it also provides more flexibility in the location of the cryostats supporting the current leads. HTS links of the type required for the accelerator technology did not exist, and significant work is being done to develop a long-length multi-conductor operating in helium gas at about 20 K.

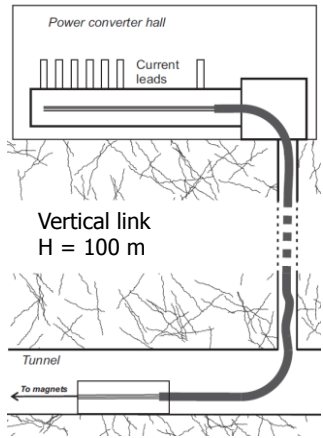

Figure 10: Schematic layout option at the LHC with the power convertors at the surface

Considerable R&D is at present being done on HTS cables for electrical utilities, and it might be a consideration that one could directly apply these technologies. However, at present this work is focused on using single or 3-phase AC conductors with high voltage insulation and liquid nitrogen cooling, and it should be noted that this is still development work yet to be concluded. Particle accelerators require high quasi- DC current carrying links with many cables (up to about 50) in parallel and cooled with liquid or gaseous helium. In the LHC there are over 50000 connecting cables with a total length of 1360 km. Thus the need specific to accelerator applications, is for a new type of link with multiple circuits, electrically isolated at around 1 kV -2 kV, carrying quasi-DC currents. The design study has to cover the options with YBCO, BSCCO and $MgB_2$ at a temperature of 20 K as well as the electrical connections between HTS and LTS.

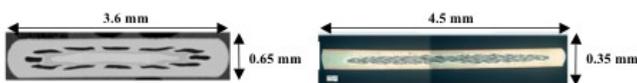

Figure 11: Two tape conductor candidates for a SC link

The task is studying the various conductors available on the market to find suitable candidates (Fig. 11) and lengths of up to 1 km of several tapes have been procured for this. Prototype cables are being tested at several partner of the collaboration. Studies and tests of the electrical joints between tapes (splices) are being done.

For the LHC applications several link types are being designed and one design case can be seen in Fig. 12. The task will conclude with the construction and test at CERN of prototype link segments.

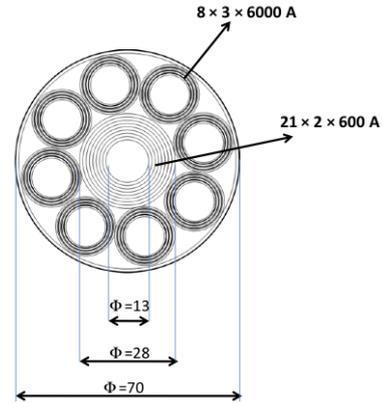

Figure 12: Example of a link layout with multiple conductors in a concentric geometry

## SHORT PERIOD HELICAL SUPERCONDUCTING UNDULATOR

The aim is to increase the achievable magnetic field level in short period undulator magnets through the use of advanced materials ($Nb_3Sn$ conductors) and innovative designs (helical coils). For example, single pass free electron lasers (e.g. X-FEL, FERMI@ELETTRA) could cover a wider wavelength range through field enhancement, or alternatively, operate at significantly lower electron energy. Additionally, short period undulator magnets could be used in the production of positrons for any future lepton collider and increased magnetic field levels will increase the positron yield and also allow for savings.

Previously an Nb-Ti helical undulator achieved an on-axis field of 0.86 T with a peak coil field of 2.74 T at 4.2 K. The aim is to reach B = 1.5 T on-axis with a peak field on the coil of 4.4 T and a period of 11.5 mm on a winding bore of 6.35 mm. $Nb_3Sn$ will be tried to get the higher current densities at the 4 T - 5 T range combined with temperature margins of several Kelvin needed in the synchrotron light environments in the accelerator.

Known challenges are a sufficiently thin $Nb_3Sn$ insulation system compatible with the heat treatment, the hoop stress in the wire and a controlled winding system for single (insulated) wires in a helical groove. First winding tests with a 0.5 mm thick wire (0.65 mm with insulation) have given encouraging results (Fig. 13). The task will design, construct and test a short (500 mm) undulator model and compare the results with the NB-Ti model previously tested.

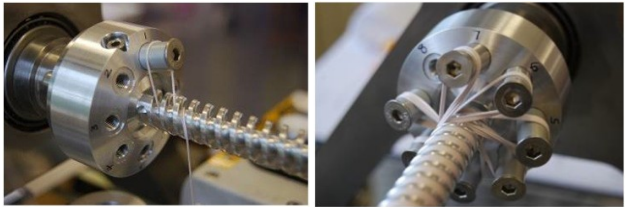

Figure 13: Winding trails for the helical undulator

## SUPPORT STUDIES

The aim of the support studies is to study radiation effects on and thermal behaviour of Nb3Sn magnets to prepare for accelerator application of these magnets. For the EuCARD-Fresca2 magnet of task2 solutions for the insulation and the thermal design are to be proposed possibly compatible with accelerator applications.

Magnets in accelerators like the upgraded LHC and are subjected to very high radiation doses. In the low beta insertion quadrupole the integrated peak dose on the coil can attain 50 MGy over the lifetime of the HL-LHC. The electrical insulation employed on the coils need to be resistant to this radiation. A certification program for the radiation resistance is needed in parallel to the modelling efforts for such magnets. The same radiation is also depositing heat in the coils. The heat removal from the coils needs to be modelled. These models have to be supported with measurements. A thermal design of the dipole model coil can then be made.

Four potential impregnation materials will be tested (RAL mix 71, Epoxy TGPAP-DDS(2002), LARP CTD101K with filler ceramic and 3 Cyanite Ester mixes) to assess their suitability for high radiation environments. For this mechanical, electrical and thermal conductivity measurements will be done on samples irradiated with and electron beam up to 50 MGy. The irradiation will be done at IJP Swierk (Po) in 2011.

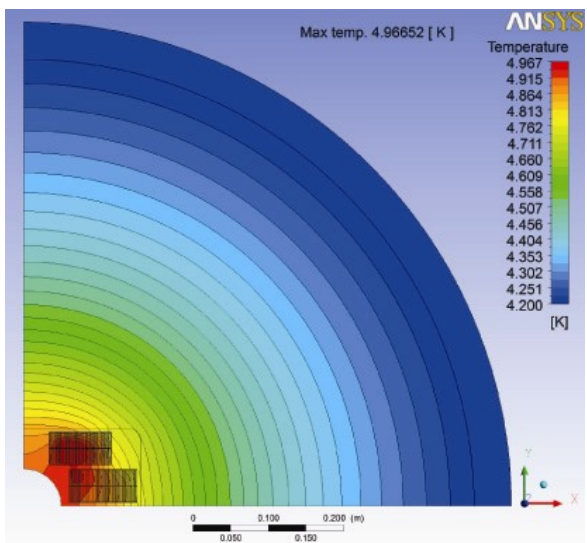

Figure 14: Calculated temperature distribution in the magnet at a total heat load of 0.167 W/m during ramping (start temperature 4.2 K)

Thermal models of $Nb_3Sn$ magnets are being used to study cool-down scenarios and steady state heat load (at 4.2 K and 1.9 K) on the coils. In Fig 14 a thermal map from a preliminary steady state heat load study can be found.

## FUTURE R&D

At present ESGARD has launched preparations for a successor project for EuCARD (EuCARD2), which is to start by the beginning of 2012. Four institutes (CERN, CEA, LBNL and KEK) envisage taking the lead in starting a larger collaboration to develop high field magnets for HE-LHC. Following the development of the 13 T wide aperture magnet in EuCARD and the HTS insert and under the condition that these developments are successful, the logical successor project is to prepare for a high field magnet for a HE-LHC type collider application.

The project could consist of the following R&D items:
1) Make a design study for a 20 T magnet for HE-LHC.
2) Construct a technology demonstrator model dipole magnet in the 15 T - 18 T range.
3) Conductor development for the 20 T field range.

For the LHC it took 22 years from the start of the magnet development to the switch-on of the machine. One has to start now with the development of 20 T magnets in order to be ready for HE-LHC in the 20+ year time scale. Experience from the LHC and presently from LARP, with the development of the low beta insertion quadrupoles for HL-LHC, indicates that this has to be done in a large international collaboration.

## ACKNOWLEDGMENT

I would like to thank the colleagues of the FP7-EuCARD-HFM collaboration for their contributions to this paper.